\documentclass [12 pt]{article}
\usepackage{psfig}
\usepackage{epsfig}
\begin{document}
\title{Island Resonances}
\author{George Parzen}
\date{\today  }
\maketitle
\begin{abstract}
Resonances driven by the error field multipoles in the magnets
often show themselves as 
islands in phase plots of the particle motion.
In order to be able to measure the strength of the 
island resonance, and then to correct it, it is helpful to study how these 
resonances show themselves.
In particular , one can study how the presence 
of the island resonances distort the tune dependence on the amplitude,
and the emittance growth caused by the island resonance.
 Islands in 
4-dimensional phase space, unlike islands in 2-dimensional phase space, 
are not easy to visualize. This study will  show 
how to visualize the islands in 4-dimensional phase space by studying the 
tune dependence on the amplitude of the particle.
These effects are to 
some extent measureable, and can lead to a way to correct the resonance.
\end{abstract}

\section*{Introduction  }
Resonances driven by the error field multipoles often show themselves as 
islands in phase plots of the particle motion. This happens when there is 
another non-linear field present, besides the error field multipoles,
 which produces a strong enough dependence of the tune on the amplitude of 
the particle motion. Non-linear fields that produce a dependence of the tune
on the ampliude can be the quadrupole fringe field , space charge fields and 
octupole correctors. In some cases it may be advantageous to introduce a
non-linear field that produces a dependence of the tune on the amplitude of 
the particle motion. In order to be able to measure the strength of the 
island resonance, and then to correct it, it is helpful to study how these 
resonances show themselves. In particular , one can study how the presence 
of the island resonances distort the tune dependence on the amplitude
and the emittance growth caused by the island resonance. 
These effects are to 
some extent measureable, and can lead to a way to correct the resonance.
 Islands in 
4-dimensional phase space, unlike islands in 2-dimensional phase space, 
are not easy to visualize. This study will  show 
how to visualize the islands in 4-dimensional phase space by studying the 
tune dependence on the amplitude of the particle. Again, this effect is to 
some extent measureable, and can lead to a way to measure the strength of the
resonance.

\begin{figure}[tbp]
\centering
\epsfig{file=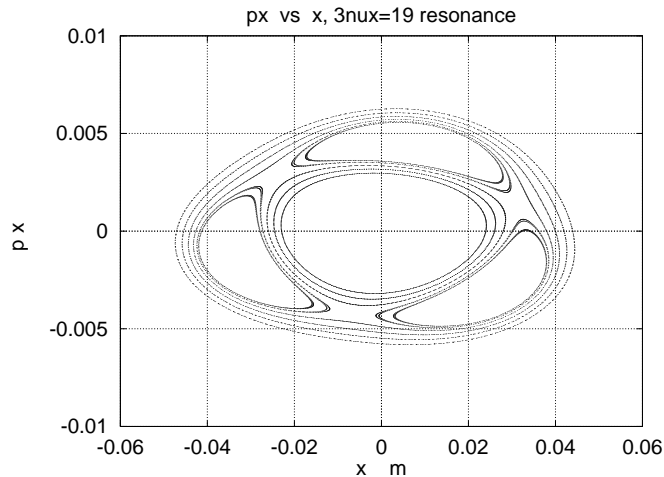,width=9.0cm}
\caption{$p_x $ vs. $x$ for the $3\nu_x=19$ resonance excited by a random $b_2$
in the SNS magnets. The initial coordinates have $x_0=24 mm$ to $x_0=44 mm$
in steps of 2 mm with $p_{x0}=0$ and $\epsilon_y=0$ .$\nu_{x0}=6.3267, \nu_{y0}=6.2267$.
In the figure $x0,px0,y0,py0,epx0,ept0,nux0,nuy0$ represent
$x_0,p_{x0},y_0,p_{y0},\epsilon_{x0},\epsilon_{y0},\epsilon_{t0},
\nu_{x0},\nu_{y0}$. }
\label{fig1}
\end{figure}

\begin{figure}[tbp]
\centering
\epsfig{file=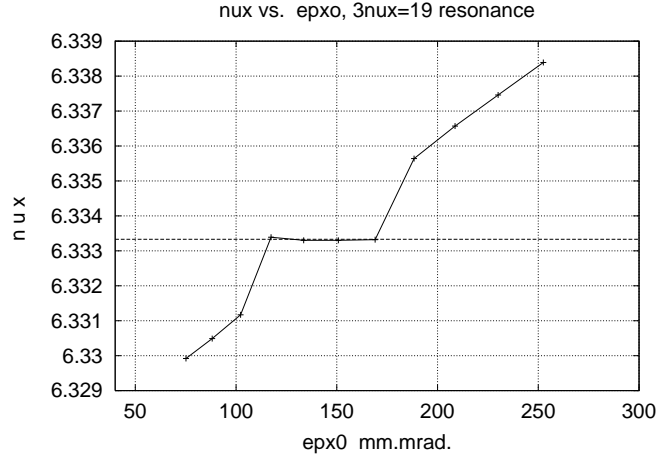,width=9.0cm}
\caption{$\nu_x$  versus $\epsilon_{x0}$ for the $3\nu_x=19$ resonance excited by a random $b_2$
in the SNS magnets. The initial coordinates have $x_0=24 mm$ to $x_0=44 mm$
in steps of 2 mm with $p_{x0}=0$ and $\epsilon_{y0}=0$ .$\nu_{x0}=6.3267, \nu_{y0}=6.2267$.
In the figure $x0,px0,y0,py0,epx0,ept0,nux0,nuy0$ represent
$x_0,p_{x0},y_0,p_{y0},\epsilon_{x0},\epsilon_{y0},\epsilon_{t0},
\nu_{x0},\nu_{y0}$.}
\label{fig2}
\end{figure}

\begin{figure}[tbp]
\centering
\epsfig{file=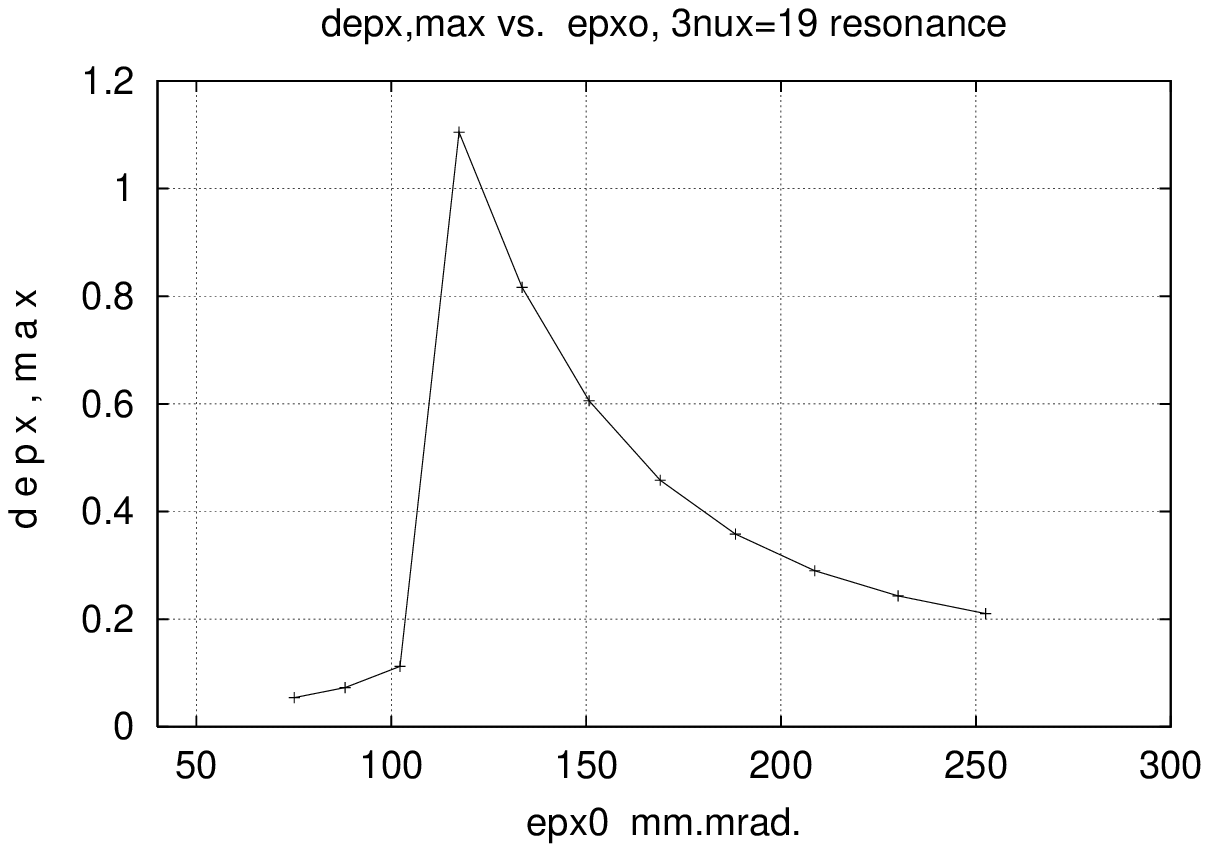,width=9.0cm}
\caption{$d\epsilon_{x,max}$ vs $\epsilon_{x0}$ for the $3\nu_x=19$ resonance 
excited by a random $b_2$ in the SNS magnets. 
The initial coordinates $x_0, p_{x0}$ 
lie along the direction in $x_0, p_{x0}$ phase space given by $p_{x0}=0$ 
and $\epsilon_y=0$. $\nu_{x0}=6.3267, \nu_{y0}=6.2267$.  
In the figure $x0,px0,y0,py0,epx0,ept0,nux0,nuy0$ represent
$x_0,p_{x0},y_0,p_{y0},\epsilon_{x0},\epsilon_{y0},\epsilon_{y0},\epsilon_{t0},
\nu_{x0},\nu_{y0}$. }
\label{fig3}
\end{figure}

\begin{figure}[tbp]
\centering
\epsfig{file=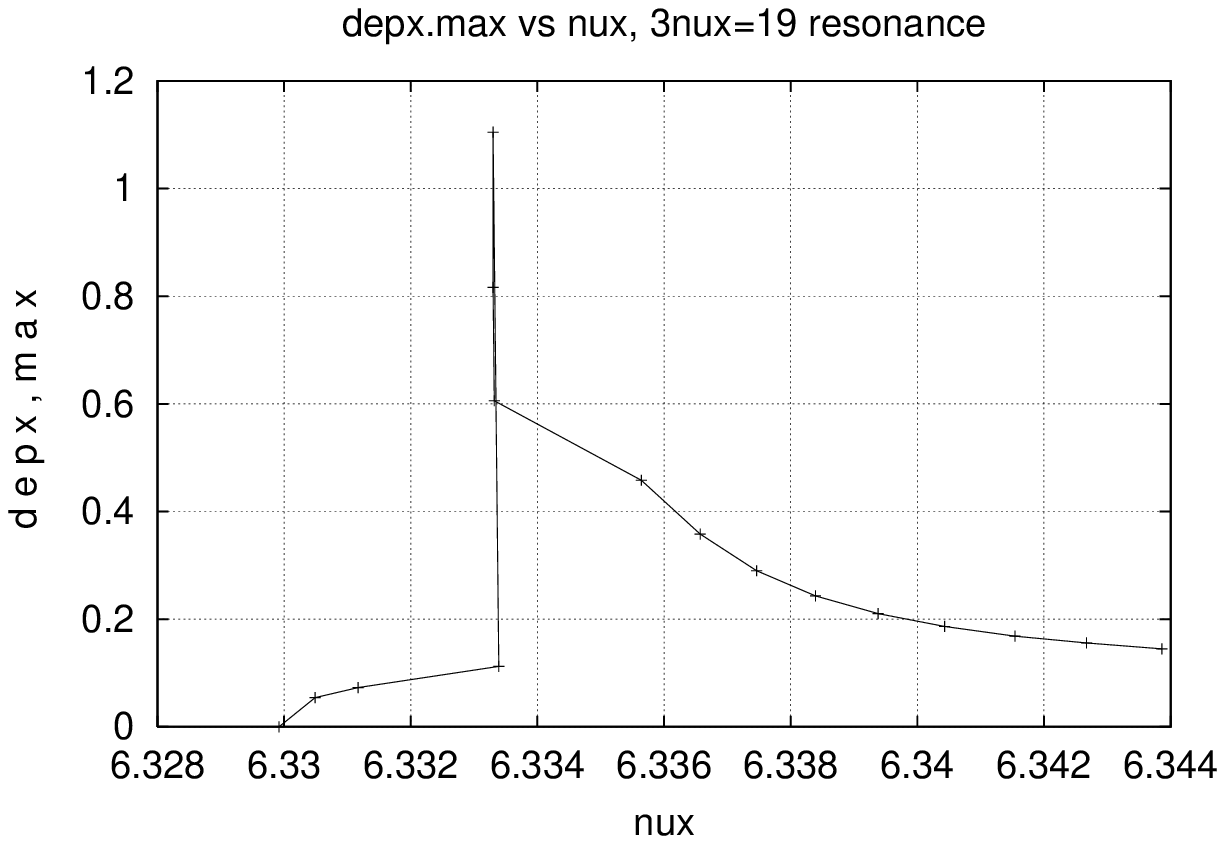,width=9.0cm}
\caption{$d\epsilon_{x,max}$ vs $\nu_x$  for the $3\nu_x=19$ resonance 
excited by a random $b_2$ in the SNS magnets. 
The initial coordinates $x_0, p_{x0}$ 
lie along the direction in $x_0, p_{x0}$ phase space given by $p_{x0}=0$ 
and $\epsilon_{y0}=0$. $\nu_{x0}=6.3267, \nu_{y0}=6.2267$.
In the figure x0,px0,y0,py0,epx0,ept0,nux0,nuy0 represent
$x_0,p_{x0},y_0,p_{y0},\epsilon_{x0},\epsilon_{y0},\epsilon_{t0},
\nu_{x0},\nu_{y0}$.}
\label{fig3b}
\end{figure}

\section*{Resonances in 2-dimenional phase space}
This section will study an island resonance in 2-dimensional phase space,
the $3 \nu_x=19$ resonance in the storage ring of the
 SNS (Spallation Neutron Source at Oak Ridge). 
This resonance 
is driven by the  random error sextupole, $b_2$, in the magnets. In the absence 
of space charge forces, a non-linear field that produces an appreciable 
tune dependence on the amplitude of the particle motion is provided by 
the fringe field of the quadrupoles \cite{ypda}, which causes the tune 
to increase with the amplitude . Starting with a zero amplitude tune of
$\nu_x=6.3267$ , $3 \nu_x=19-.02$, one finds the phase space plot 
shown in Fig.~\ref{fig1}.
This plot was generated starting with $x_0=24 mm$, $p_{x_0}=0$ 
and then increasing $x_0$
in steps of 2 mm keeping $p_{x_0}=0$ and tracking the particle for
1000 turns for each $x_0$.
The tune starts below the resonance, and increases with amplitude. 
The islands appear when the tune reaches $\nu_x=6.3333$.
The islands are reached when the 
initial amplitude is increased to $x_0=30 mm$, $p_{x_0}=0$, 
which corresponds to an initial 
horizontal emittance of about $\epsilon_{x0}=115$ mm.mrad. 

The distortion of the tune dependence on 
amplitude is shown in  Fig.~\ref{fig2} where 
the horizontal tune is plotted against $\epsilon_{x0}$.
Fig.~\ref{fig2} is generated by doing a search
along the $p_{x_0}=0$ direction, increasing $x_0$ in  steps of 2 mm. 
Fig.~\ref{fig2} shows a flat region where $\nu_x=6.3333$ 
which will be seen to indicate 
the region where $x_0,p_{x_0}$ is crossing the island. When $x_0,p_{x_0}$
 are inside the island,
the paticle motion is doing a slow oscilation around the fixed point at the 
center of the island. The particle motion then contains more than one tune, 
and the tune with the largest ampltude is the tune of the fixed point, 6.3333.
Inside the island, the tune shown in Fig.~\ref{fig2}, is this tune with 
the largest amplitude.
One can verify that the beginning and the end of the flat region occur
at the same $\epsilon_{x0}$ which correpond to 
the borders of the islands along the $p_{x_0}=0$
direction. 
Fig.~\ref{fig2} can be used to find
the width of the island along the $p_{x_0}=0$ direction which is given by the
width of the flat region in  Fig.~\ref{fig2}.

The islands generated by the $3\nu_x=19$ resonance indicate a 
growth in the particle
emittance. If one starts a particle inside the islands with the initial emittance
$\epsilon_{x0}$, then as the particle moves around the fixed point, 
its emittance will change  and
reach the maximun value of $\epsilon_{x,max}$. One can use as 
a measure of the emittance growth the quantity 
\[ d\epsilon_{x,max}=(\epsilon_{x,max}-\epsilon_{x0})/\epsilon_{x0} \]. 
$d\epsilon_{x,max}$ is plotted against
$\epsilon_{x0}$ in Fig.~\ref{fig3}. It shows a maximun value for 
$d\epsilon_{x,max}$ 
which can be used as a  measure of the strength of the resonance. 

The results shown in Fig.~\ref{fig2} and Fig.~\ref{fig3} can 
to some extent be measured experimentally
to find a way to correct this resonance with sextupole correctors.
An interesting plot that may corrrespond more closely to something that might 
be measured is to plot the emittance growth, $d\epsilon_{x,max}$ vs 
the tune, $\nu_x$. 
This plot can be found by combining Fig.~\ref{fig2} and Fig.~\ref{fig3} 
and is shown 
in Fig.~\ref{fig3b}. In this plot, there is a peak in the emittance growth 
which occurs when
$\nu_x$=6.3333, the resonance tune. The peak could be used as a measure of the 
resonance strength to correct the resonance. 

\begin{figure}[tbp]
\centering
\epsfig{file=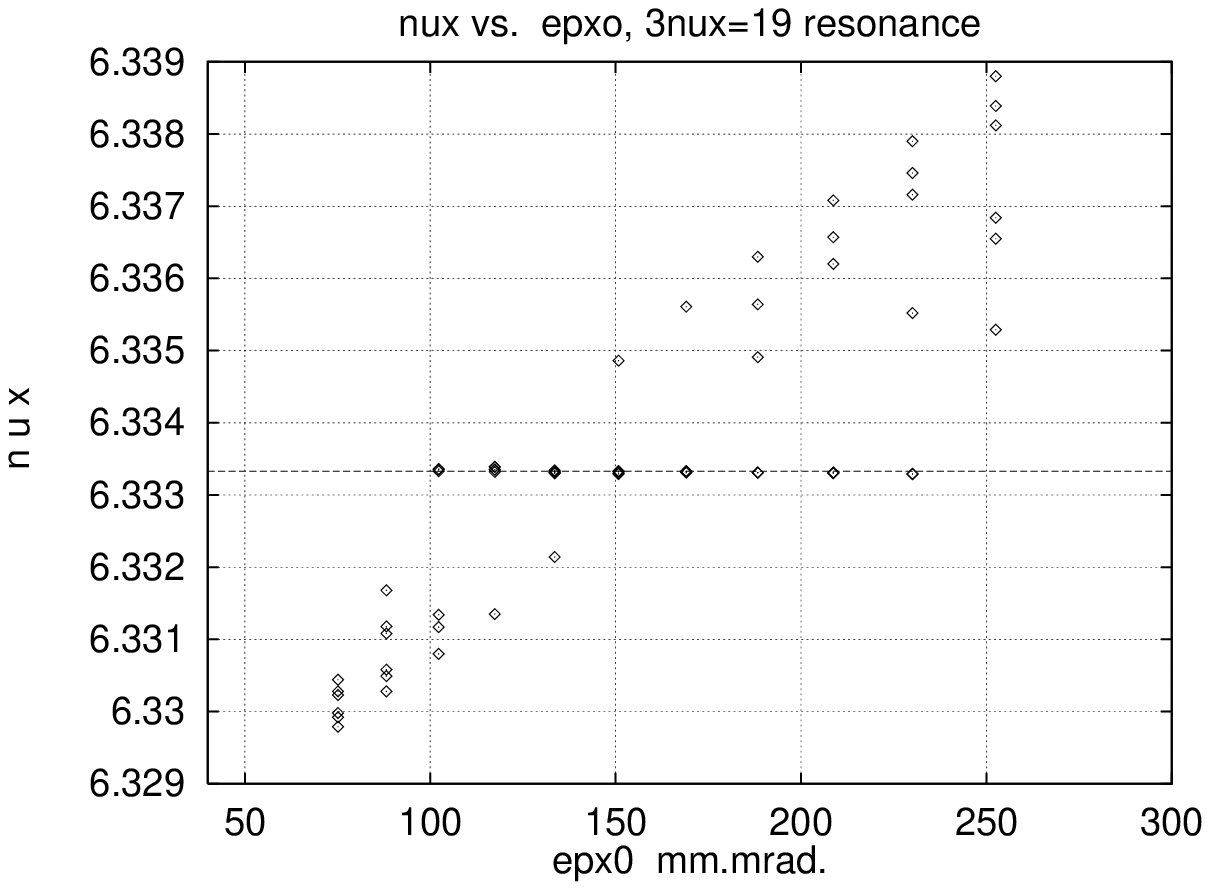,width=9.0cm}
\caption{$\nu_x$ vs $\epsilon_{x0}$ for the $3\nu_x=19$ resonance 
excited by a random b2
in the SNS magnets. The initial coordinates $x_0, p_{x0}$ lie along 6 directions in
the $x_0, p_{x0}$ phase space and $\epsilon_{y0}=0$ .$\nu_{x0}=6.3267, \nu_{y0}=6.2267$.
In the figure x0,px0,y0,py0,epx0,ept0,nux0,nuy0 represent
$x_0,p_{x0},y_0,p_{y0},\epsilon_{x0},\epsilon_{y0},\epsilon_{t0},
\nu_{x0},\nu_{y0}$.}
\label{fig4}
\end{figure}

\begin{figure}[tbp]
\centering
\epsfig{file=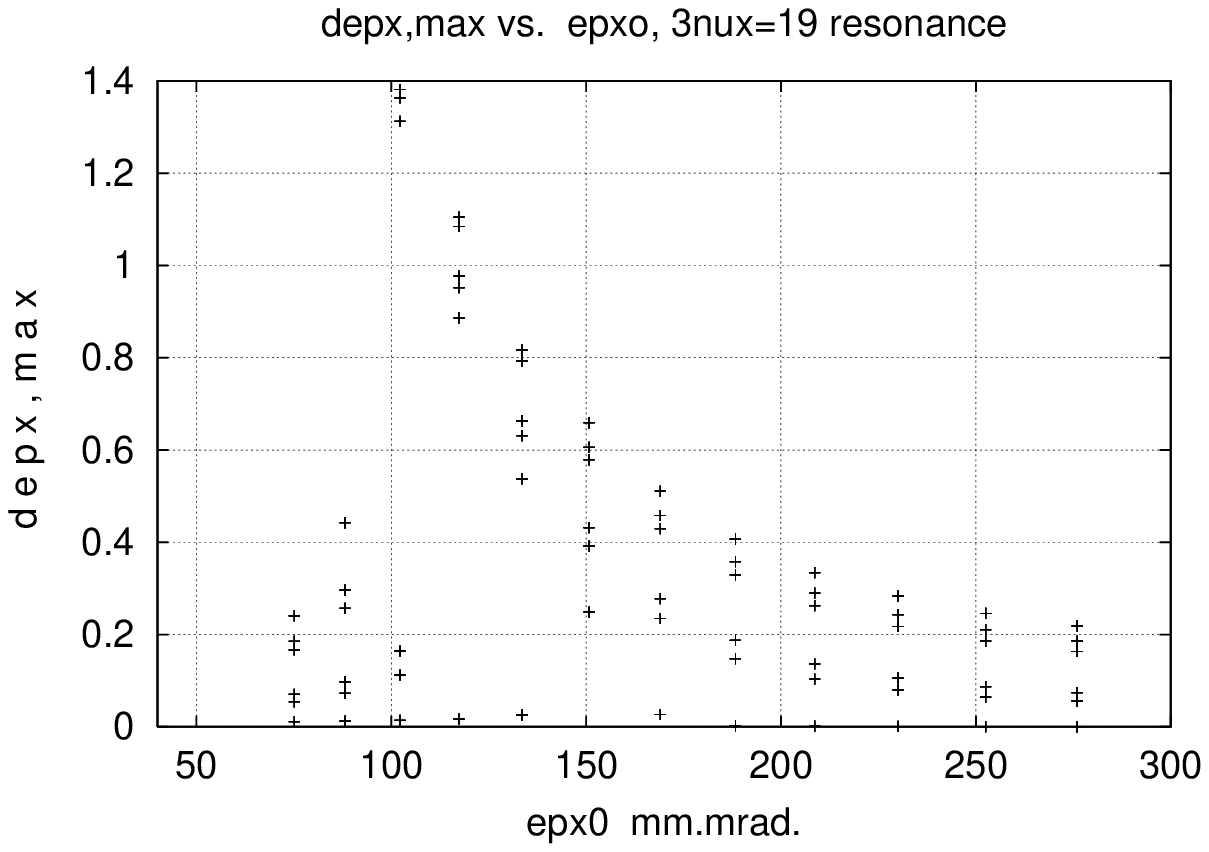,width=9.0cm}
\caption{$d\epsilon_{x,max}$ vs $\epsilon_{x0}$ for the $3\nu_x=19$ resonance excited by a random $b_2$
in the SNS magnets. The initial coordinates $x_0, p_{x0}$ lie along 6 directions in
the $x_0, p_{x0}$ phase space and $\epsilon_{y0}=0$. $\nu_{x0}=6.3267, \nu_{y0}=6.2267$.
In the figure x0,px0,y0,py0,epx0,ept0,nux0,nuy0 represent
$x_0,p_{x0},y_0,p_{y0},\epsilon_{x0},\epsilon_{y0},\epsilon_{t0},
\nu_{x0},\nu_{y0}$.}
\label{fig5}
\end{figure}

\begin{figure}[tbp]
\centering
\epsfig{file=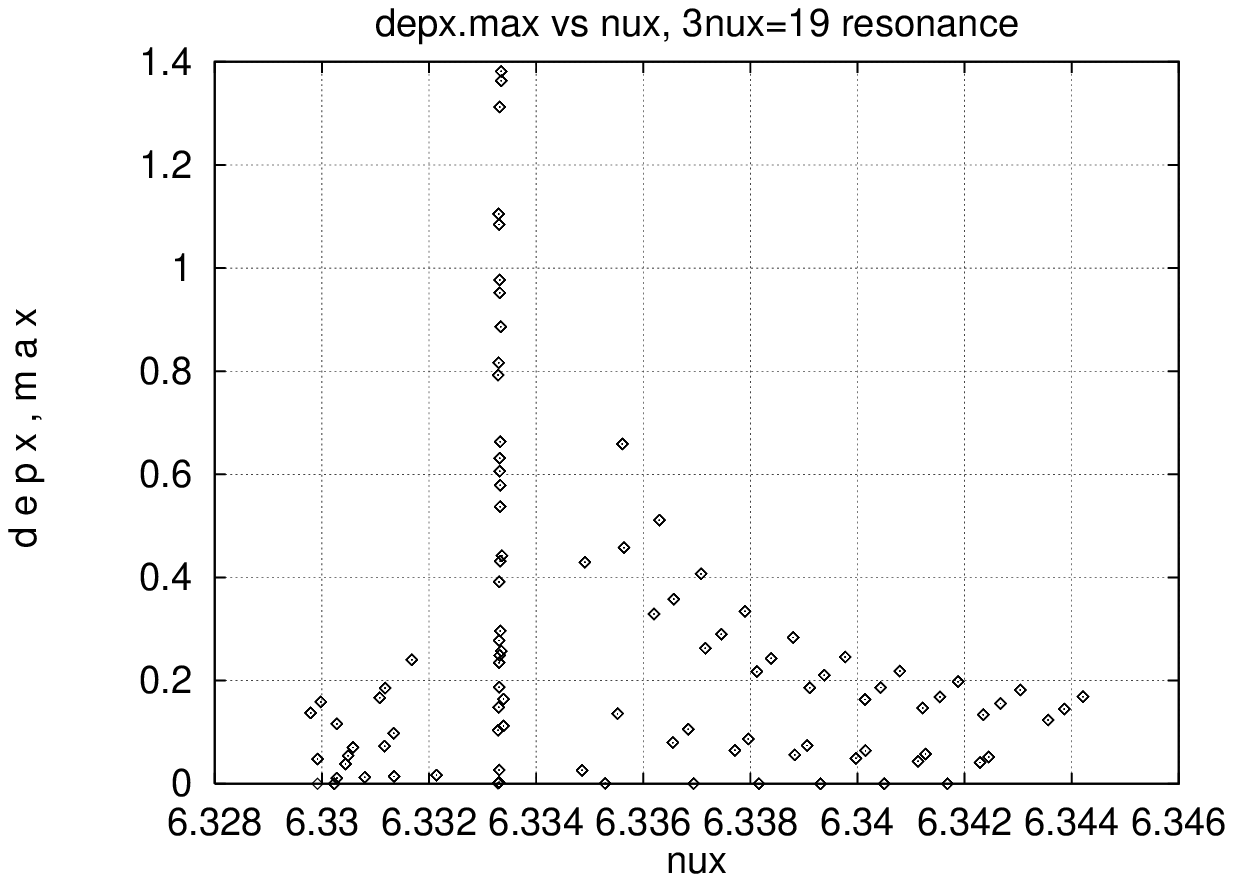,width=9.0cm}
\caption{$d\epsilon_{x,max}$ vs $\nu_x$  for the $3\nu_x=19$ resonance excited by a random $b_2$
in the SNS magnets. The initial coordinates $x_0, p_{x0}$ lie along 6 directions in
the $x_0, p_{x0}$ phase space and $\epsilon_{y0}=0$. $\nu_{x0}=6.3267, \nu_{y0}=6.2267$.
In the figure x0,px0,y0,py0,epx0,ept0,nux0,nuy0 represent
$x_0,p_{x0},y_0,p_{y0},\epsilon_{x0},\epsilon_{y0},\epsilon_{t0},
\nu_{x0},\nu_{y0}$.}
\label{fig5b}
\end{figure}

Using the flat region in the tune dependence on amplitude along the $p_{x0}=0$ 
direction to indicate the strength of the resonance  can sometimes lead to an error
in setting the strengths of the sextupole correctors. Changing the strength of the 
correctors can cause the islands to move, and the width of the island as 
indicated by the flat region in the tune dependence on the amplitude may,
for example,  appear smaller because one is now crossing the island 
at a  place where it is narrower. Also, when the islands move the the emittance
growth , $d\epsilon_{x,max}$ may seem smaller.
In order to avoid this error due to the movement of 
the islands when the corrector strengths are changed, one has to measure the tune 
dependence on amplitude along enough different directions in phase space 
so that one finds the largest width of the island and the largest emittance growth
for the $x_0,p_{x0}$ inside the island. The same sort of argument
also applies  in finding the emittance growth dependence on the amplitude. 
In Fig.~\ref{fig4} and Fig.~\ref{fig5},
 results are given  found by 
going out along many directions
in phase space. One needs just enough directions to cover one of the 
three islands in Fig.~\ref{fig1}. Here 6 directions were used in the 
region  of phase space
covered by one island.

In Fig.~\ref{fig4} that plots $\nu_x$ versus $\epsilon_{x0}$, for each 
$\epsilon_{x0}$ there are now 6 points plotted
corresponding to the 6 directions. Points that have the tune, nux=6.3333, lie 
inside the island.The width of the island in $\epsilon_{x0}$ as given by 
the width of the points that lie on the nux=6.3333 line and is much larger than the
width found using just the $p_{x0}$=0 direction as it includes the direction 
that crosses the island close to where the island is widest. In the same way, 
Fig.~\ref{fig5}  now shows the largest emittance growth for the $x_0,p_{x0}$
that are inside the 
island.

In applying the above results to develope a procedure for using the correctors 
to correct a resonance, one will probably use a sample of the $x_0,p_{x0}$ that 
is convenient for a particular ring and its injection system. One would then
measure either the tune or the emittance growth of enough particles in the 
sample to be able to find the width of the island from the tune measurement
 or the largest emittance growth
for paticles in the sample. One has to have a large enough sample that one is not 
misled by the movement of the islands when the excitation of the correctors are 
changed. To measure the largest emittance growth
for particles in the sample, one can measure the frequency 
spectrum of the  betatron
 oscillations. Particles inside an island will have the frequency that corresponds 
to the resonant tune 6.3333. The largest amplitude found in the frequency spectrum 
for the resonant frequency, can be used as a measure of the 
largest emittance growth.
Something like this was done at RHIC \cite{VPAFFP}.
The measurements can be done for a 
weak beam  (no space charge effects) as it has been shown \cite{AFGP} that the 
resonance correction for resonances generated by magnet errors in the absence of 
space charge will also work fairly well in the presence of space charge.

An interesting plot that may corrrespond more closely to something that might 
be measured is to plot the emittance growth, $d\epsilon_{x,max}$ vs 
the tune, $\nu_x$. This plot can be found by combining Fig.~\ref{fig4} 
and Fig.~\ref{fig5} and is shown 
in Fig.~\ref{fig5b}

\begin{figure}[tbp]
\centering
\epsfig{file=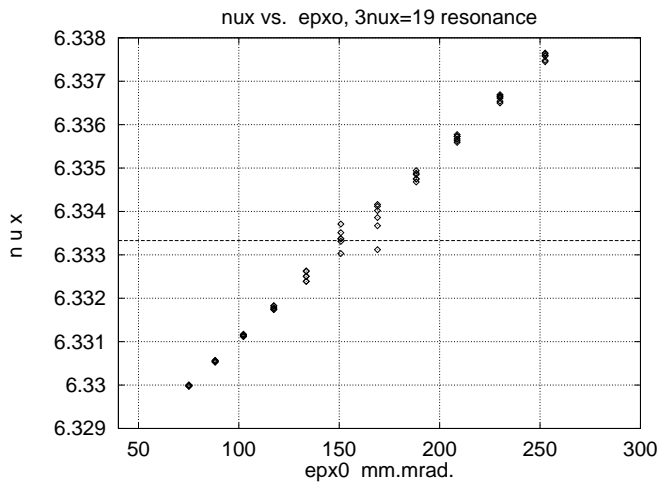,width=9.0cm}
\caption{$\nu_x$ vs $\epsilon_{x0}$ for the $3\nu_x=19$ resonance excited by a random 
$b_2$
in the SNS magnets and corrected using two sextupole correctors. 
The initial coordinates $x_0, p_{x0}$ lie along 6 directions in
the $x_0, p_{x0}$ phase space and $\epsilon_{y0}=0$. $\nu_{x0}=6.3267, \nu_{y0}=6.2267$.
In the figure x0,px0,y0,py0,epx0,ept0,nux0,nuy0 represent
$x_0,p_{x0},y_0,p_{y0},\epsilon_{x0},\epsilon_{y0},\epsilon_{t0},
\nu_{x0},\nu_{y0}$.}
\label{fig6}
\end{figure}

\begin{figure}[tbp]
\centering
\epsfig{file=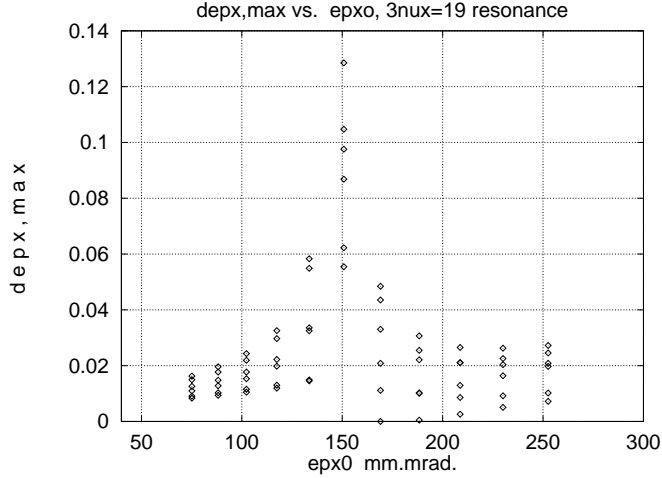,width=9.0cm}
\caption{$d\epsilon_{x,max}$ vs $\epsilon_{x0}$  for the $3\nu_x=19$ resonance excited by a random $b_2$
in the SNS magnets and corrected using two sextupole correctors. 
The initial coordinates $x_0, p_{x0}$ lie along 6 directions in
the $x_0, p_{x0}$ phase space and $\epsilon_{y0}=0$ .$\nu_{x0}=6.3267, \nu_{y0}=6.2267$.
In the figure x0,px0,y0,py0,epx0,ept0,nux0,nuy0 represent
$x_0,p_{x0},y_0,p_{y0},\epsilon_{x0},\epsilon_{y0},\epsilon_{t0},
\nu_{x0},\nu_{y0}$.}
\label{fig7}
\end{figure}

The correction of the island resonances for the $3\nu_x=19$  example can be done 
with two sextupole correctors properly located around the ring.The two correctors 
can then be adjusted one at a time, and for each setting of the correctors, 
the island width can measured from the dependnce of the tune on the amplitude as
shown in Fig.~\ref{fig4}, or the dependence of the amplitude growth on the tune
as shown by combining Fig.~\ref{fig4} and Fig.~\ref{fig5}. Simulating 
this procedure gives the results shown in Fig.~\ref{fig6} and Fig.~\ref{fig7}.
The emittance growth ,the maximun depx,max, is reduced by almost a factor of 10 by 
the best setting of the correctors that was found.

The theoretical result for the width of the island indicates that the 
width can be made zero when the correctors are set so that $d\nu_{30}$=0 where
$d\nu_{30}$ is the stopband \cite{GG} of the $3\nu_x=19$ resonance due to the  field errors in 
the magnets that are driving this resonance. According to the theoretical result 
the  required setting of the correctors does not depend on the field
producing the dependence of the tune on amplitude.
 This result is not useful for setting
the correctors as the errors in the magnets are not known. However, this 
result helps to explain why the setting of the correctors in the absence of 
space charge fields will also correct the resonance when the space charge fields
are present. It was found for the $3\nu_x=19$ resonance that setting of the correctors 
to make $d\nu_{30}$=0 will only reduce the emittance growth, the maximun $d\epsilon_{x,max}$, by a factor of 
about 3 instead of the factor of 10 found above. It was found that for the SNS
\cite{AFGP} this weaker correction that makes $d\nu_{30}$=0 was good enough to reduce the
beam  losses due to the resonance and space charge fields.

\begin{figure}[tbp]
\centering
\epsfig{file=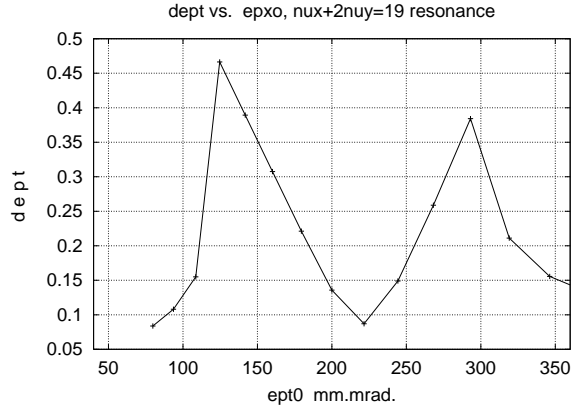,width=8.0cm}
\caption{The emittance spread $d\epsilon_t$ vs $\epsilon_{t0}$  for the $\nu_x+2\nu_y=19$ resonance 
excited by a random $b_2$
in the SNS magnets . 
The initial coordinates $x_0,p_{x0},y_0,p_{y0}$ lie along the direction in
 $x_0,p_{x0},y_0,p_{y0}$ phase space given by $p_{x0}=0$, $p_{y0}=0$ and $\epsilon_{y0}=\epsilon_{x0}$.$\nu_{x0}=6.3933$, $\nu_{y0}=6.2933$.
In the figure x0,px0,y0,py0,epx0,ept0,nux0,nuy0 represent
$x_0,p_{x0},y_0,p_{y0},\epsilon_{x0},\epsilon_{y0},\epsilon_{t0},
\nu_{x0},\nu_{y0}$.}
\label{fig3-1}
\end{figure}

\section*{Resonances in 4-dimenional phase space}

\begin{figure}[tbp]
\centering
\epsfig{file=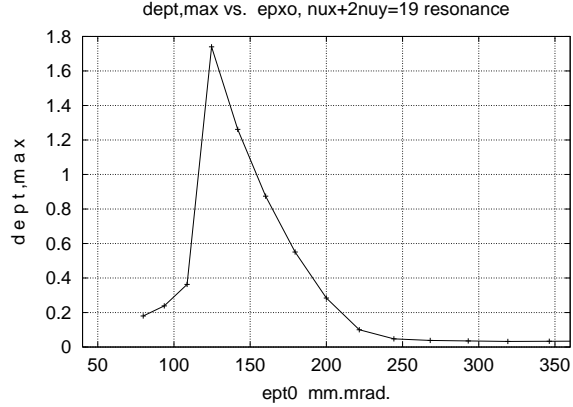,width=8.0cm}
\caption{The emittance sgrowth $d\epsilon_{t,max}$  vs $\epsilon_{t0}$  for the $\nu_x+2\nu_y=19$ resonance excited by a random $b_2$
in the SNS magnets . 
The initial coordinates $x_0,p_{x0},y_0,p_{y0}$ lie along the direction in
 $x_0,p_{x0},y_0,p_{y0}$ phase space given by $p_{x0}=0$, $p_{y0}=0$ and $\epsilon_{y0}=\epsilon_{x0}$.$\nu_{x0}=6.3933$, $\nu_{y0}=6.2933$.
In the figure x0,px0,y0,py0,epx0,ept0,nux0,nuy0 represent
$x_0,p_{x0},y_0,p_{y0},\epsilon_{x0},\epsilon_{y0},\epsilon_{t0},
\nu_{x0},\nu_{y0}$.}
\label{fig3-2}
\end{figure}

\begin{figure}[tbp]
\centering
\epsfig{file=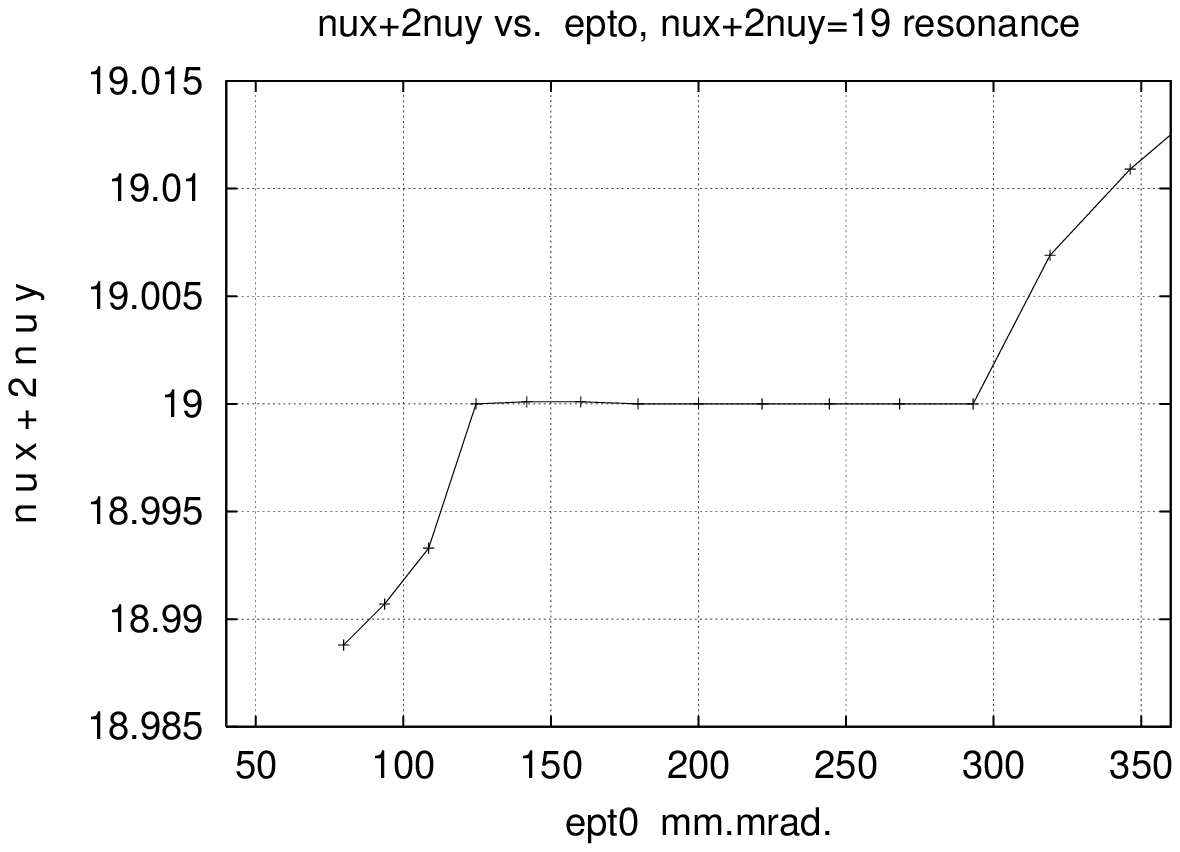,width=8.0cm}
\caption{$\nu_x+2\nu_y$  vs $\epsilon_{t0}$  for the $\nu_x+2\nu_y=19$ resonance 
excited by a random $b_2$
in the SNS magnets . 
The initial coordinates $x_0,p_{x0},y_0,p_{y0}$ lie along the direction in
 $x_0,p_{x0},y_0,p_{y0}$ phase space given by $p_{x0}=0$, $p_{y0}=0$ and $\epsilon_{y0}=\epsilon_{x0}$.$\nu_{x0}=6.3933$, $\nu_{y0}=6.2933$.
In the figure x0,px0,y0,py0,epx0,ept0,nux0,nuy0 represent
$x_0,p_{x0},y_0,p_{y0},\epsilon_{x0},\epsilon_{y0},\epsilon_{t0},
\nu_{x0},\nu_{y0}$.}
\label{fig3-3}
\end{figure}

This section will study an island resonance in 4-dimensional phase space,
the $\nu_x+2\nu_y=19$ in the storage ring of the SNS (Spallation Neutron Source 
at Oak Ridge). 
This resonance 
is driven by the  random error sextupole, $b_2$, in the magnets. Unlike, the
$3\nu_x=19$ resonance in 2-dimensional phase space, it is difficult to visualise
the islands associated with the $\nu_x+2\nu_y=19$ resonance by looking at 
the particle motion in 4-dimensional phase space. Instead, we will look at the
distortion due to the resonance on the dependence of the tune and the emittance
growth on the amplitude of the particle motion. In Fig.~\ref{fig3-1},
the spread
in total emittance ,
    \[ d\epsilon_{t}=(\epsilon_{t,max}-\epsilon_{t,min})/(\epsilon_{t,max}+\epsilon_{t,min})\]
 is plotted against 
the initial total emittance, $\epsilon_{t0}$. $\epsilon_{t,max}$ 
is the largest total emittance and
$\epsilon_{t,min}$ is the smallest total emittance reached by particle 
with the initial coordinates, $x_0,p_{x0},y_0,p_{y0}$, 
and initial total emittance, $\epsilon_{t0}$,tracked for 1000 turns.
Points were found along a particular direction in phase space given by 
$p_{x0}=0$, $p_{y0}=0$ and $\epsilon_{x}=\epsilon_{y}$.The 
emittance spread becomes large in the region
$\epsilon_{t0}$=125 to $\epsilon_{t0}$=300, and this region may be 
used as a measure of the width 
of the island along this particular direction , $p_{x0}=0$, $p_{y0}=0$ 
and $\epsilon_{x}=\epsilon_{y}$. The low 
value of $d\epsilon_{t}$ near the center of region shows that 
this particular direction
goes close to the center of the island where we expect $d\epsilon_{t}$=0.
Fig.~\ref{fig3-2} shows the emittance growth, 
\[  d\epsilon_{t,max}=(\epsilon_{t,max}-\epsilon_{t0})/\epsilon_{t0}  \] 
as a function of
$\epsilon_{t0}$. The goal of a correction scheme would be to 
reduce the maximun value
of $d\epsilon_{t,max}$ in ths plot. In the 2-dimensional case 
of the $3\nu_x=19$ resonance, 
the plot of $\nu_x$ vs. $\epsilon_{x0}$ had a flat region where $\nu_x$ 
had the constant value
of 6.3333 which  gave the width of the island. It was found for the 
$\nu_x+2\nu_y =19$ resonance, a flat region would be found if we plotted
$\nu_x+2\nu_y $ vs. $\epsilon_{t0}$ and the constant value in the 
flat region is $\nu_x+2\nu_y =19$.
This is shown in Fig.~\ref{fig3-3}. The width of the flat 
region coincides with the 
width of the island as found from Fig.~\ref{fig3-1}, 
which plots the emittance spread vs. $\epsilon_{t0}$. 
For coupled motion, the particle motion 
contains more than one tune, and what is plotted in 
Fig.~\ref{fig3-3} is the main tune 
or the tune with the largest amplitude.

\begin{figure}[tbp]
\centering
\epsfig{file=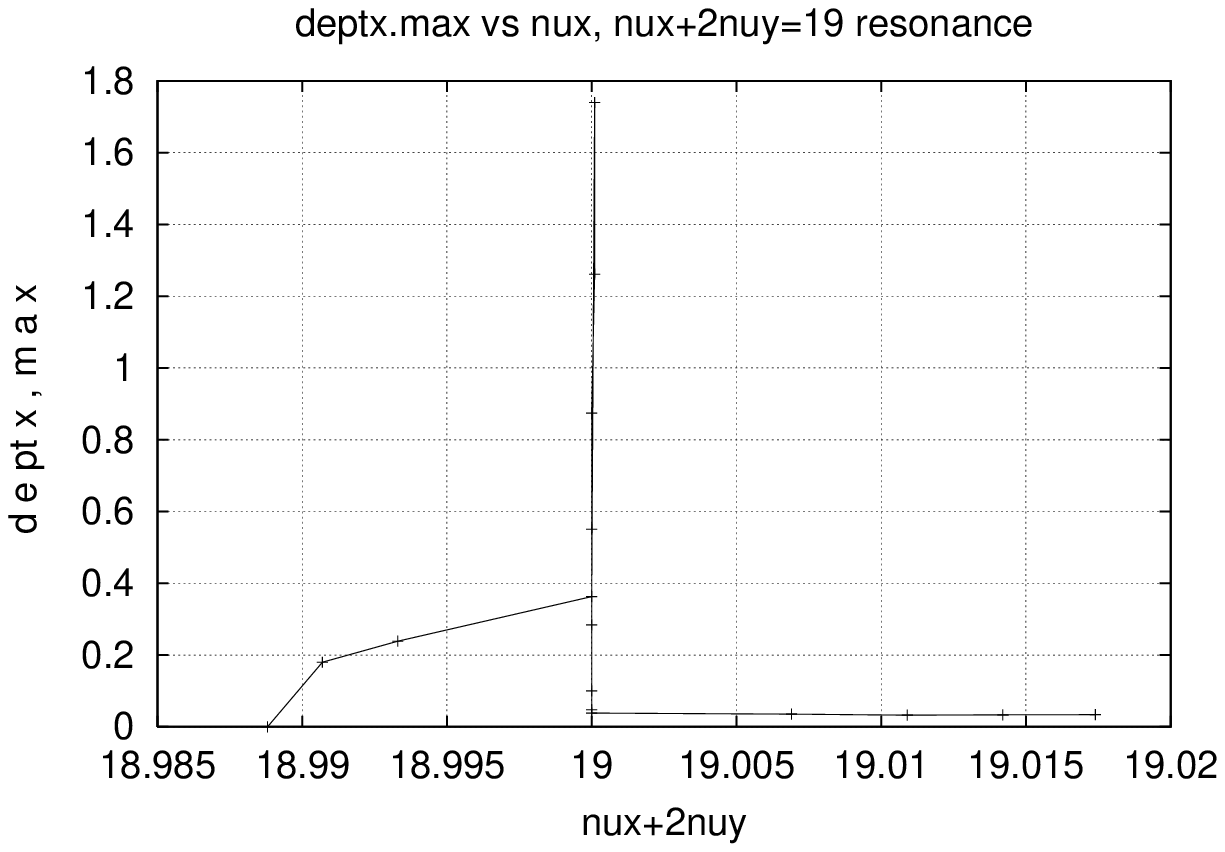,width=8.0cm}
\caption{$d\epsilon_{t,max}$ vs $\nu_x+2\nu_y$  for the $\nu_x+2\nu_y=19$ resonance 
excited by a random $b_2$
in the SNS magnets . 
The initial coordinates $x_0,p_{x0},y_0,p_{y0}$ lie along the direction in
 $x_0,p_{x0},y_0,p_{y0}$ phase space given by $p_{x0}=0$, $p_{y0}=0$ and $\epsilon_{y0}=\epsilon_{x0}$.$\nu_{x0}=6.3933$, $\nu_{y0}=6.2933$.
In the figure x0,px0,y0,py0,epx0,ept0,nux0,nuy0 represent
$x_0,p_{x0},y_0,p_{y0},\epsilon_{x0},\epsilon_{y0},\epsilon_{t0},
\nu_{x0},\nu_{y0}$.}
\label{fig3-3b}
\end{figure}

\begin{figure}[tbp]
\centering
\epsfig{file=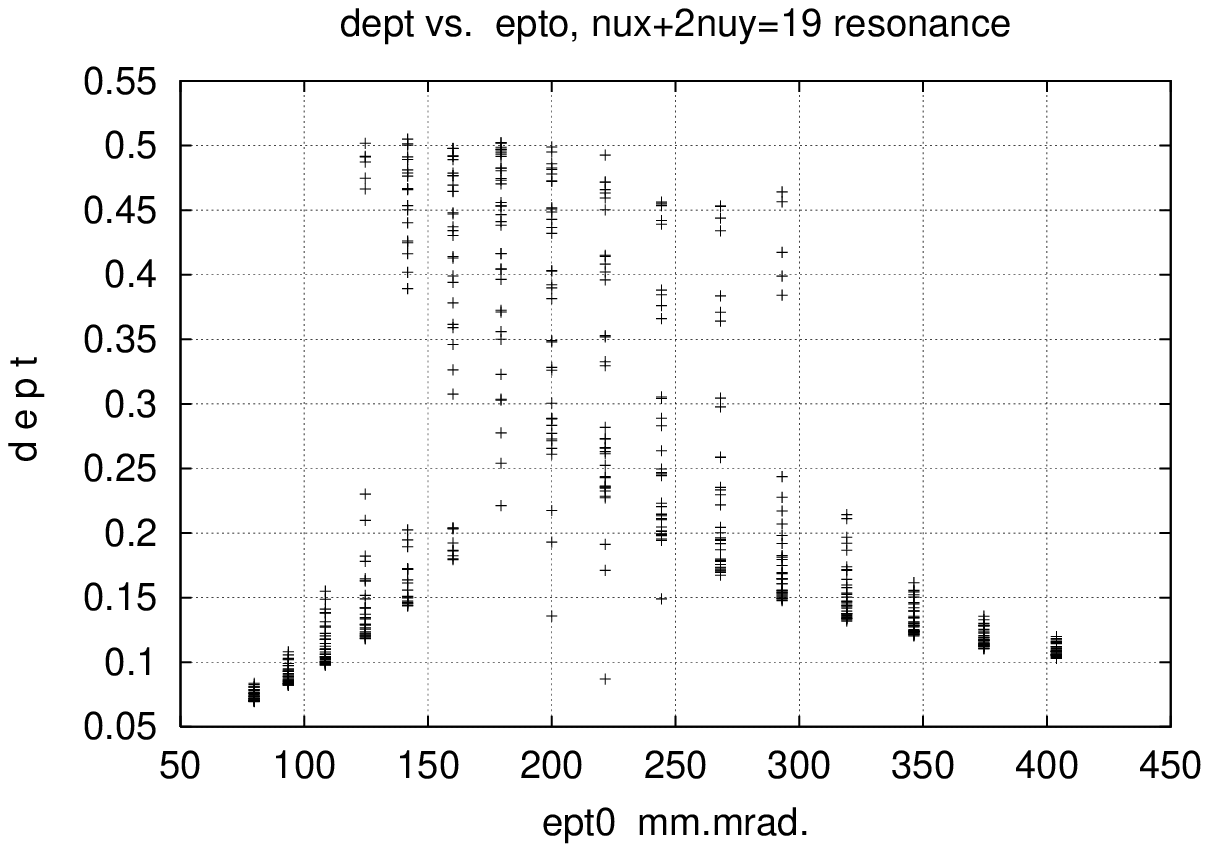,width=9.0cm}
\caption{The emittance spread $d\epsilon_t$  vs $\epsilon_{t0}$  for the $\nu_x+2\nu_y=19$ resonance excited by a random $b_2$
in the SNS magnets .
The initial coordinates $x_0,p_{x0},y_0,p_{y0}$ lie along 6 directions in
the $x_0,p_{x0}$ phase space and 6 directions in $y_0,p_{y0}$ space . $\nu_{x0}=6.3933$, $\nu_{y0}=6.2933$. 
In the figure x0,px0,y0,py0,epx0,ept0,nux0,nuy0 represent
$x_0,p_{x0},y_0,p_{y0},\epsilon_{x0},\epsilon_{y0},\epsilon_{t0},\nu_{x0},\nu_{y0}$.}
\label{fig3-4}
\end{figure}

\begin{figure}[tbp]
\centering
\epsfig{file=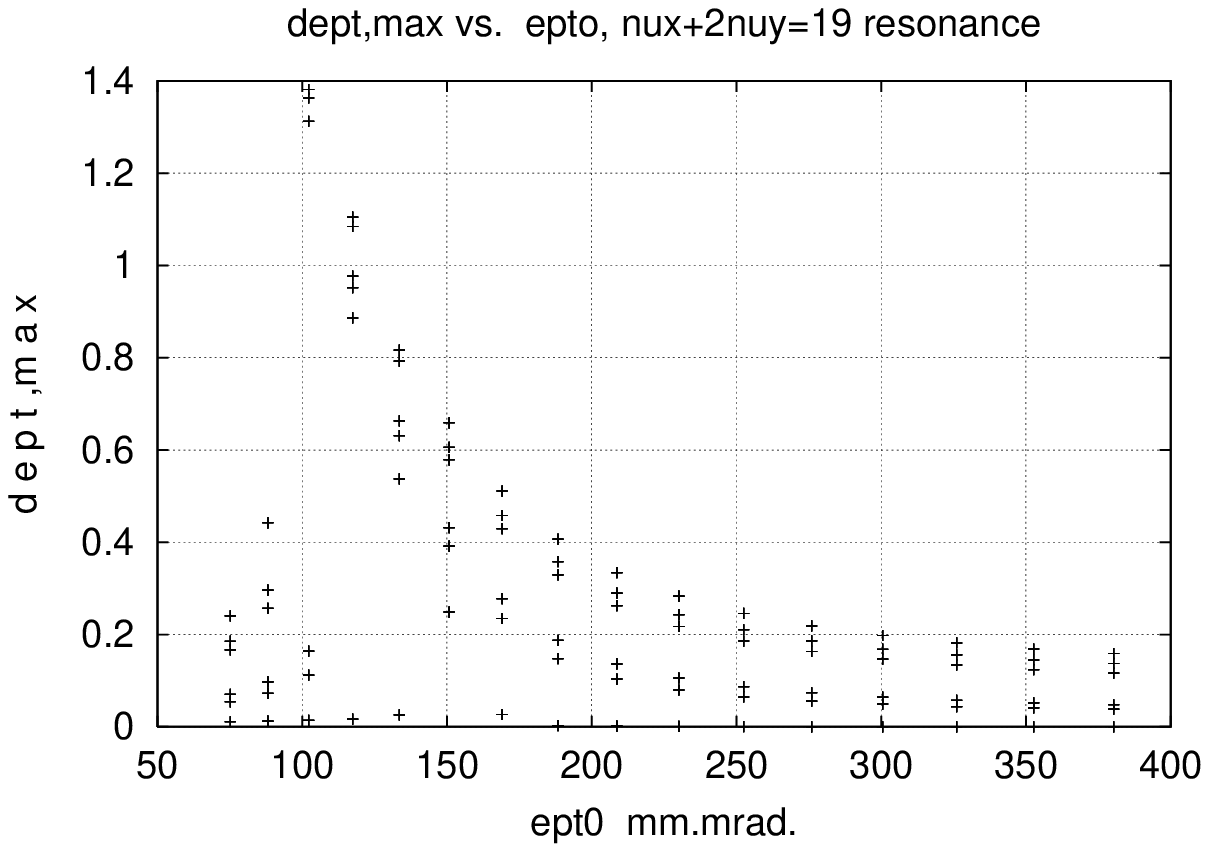,width=8.0cm}
\caption{The emittance sgrowth $d\epsilon_{t,max}$  vs $\epsilon_{t0}$  for the $\nu_x+2\nu_y=19$ resonance excited by a random $b_2$
in the SNS magnets .
he initial coordinates $x_0,p_{x0},y_0,p_{y0}$ lie along 6 directions in
the $x_0,p_{x0}$ phase space and 6 directions in $y_0,p_{y0}$ space . 
$\nu_{x0}=6.3933$, $\nu_{y0}=6.2933$.
In the figure x0,px0,y0,py0,epx0,ept0,nux0,nuy0 represent
$x_0,p_{x0},y_0,p_{y0},\epsilon_{x0},\epsilon_{y0},\epsilon_{t0},
\nu_{x0},\nu_{y0}$.} 
\label{fig3-5}
\end{figure}

\begin{figure}[tbp]
\centering
\epsfig{file=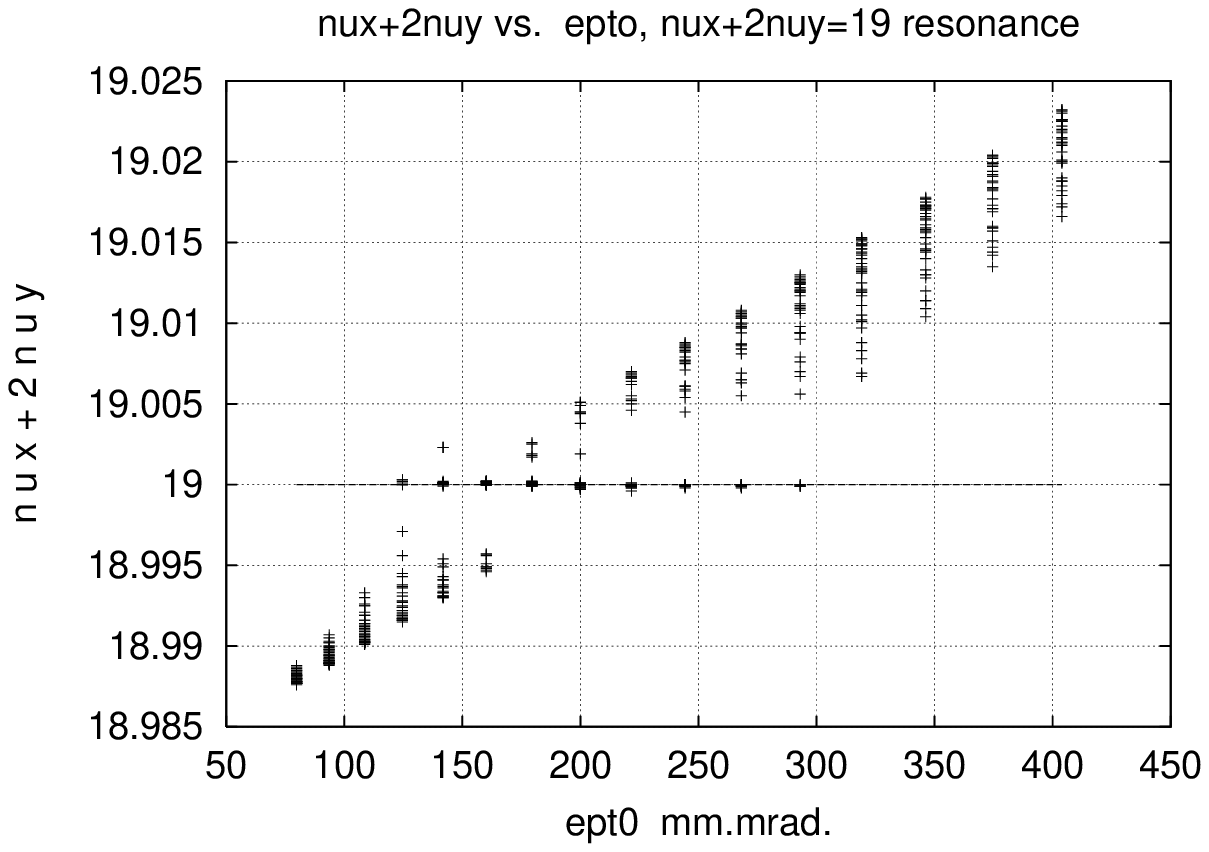,width=8.0cm}
\caption{$\nu_x+2\nu_y$  vs $\epsilon_{t0}$ for the $\nu_x+2\nu_y=19$ resonance 
excited by a random $b_2$
in the SNS magnets . 
The initial coordinates $x_0,p_{x0},y_0,p_{y0}$ lie along 6 directions in
the $x_0,p_{x0}$ phase space and 6 directions in y0,py0 space .$\nu_{x0}=6.3933$,$\nu_{y0}=6.2933$.
In the figure x0,px0,y0,py0,epx0,ept0,nux0,nuy0 represent
$x_0,p_{x0},y_0,p_{y0},\epsilon_{x0},\epsilon_{y0},\epsilon_{t0},
\nu_{x0},\nu_{y0}$.}
\label{fig3-6}
\end{figure}

\begin{figure}[tbp]
\centering
\epsfig{file=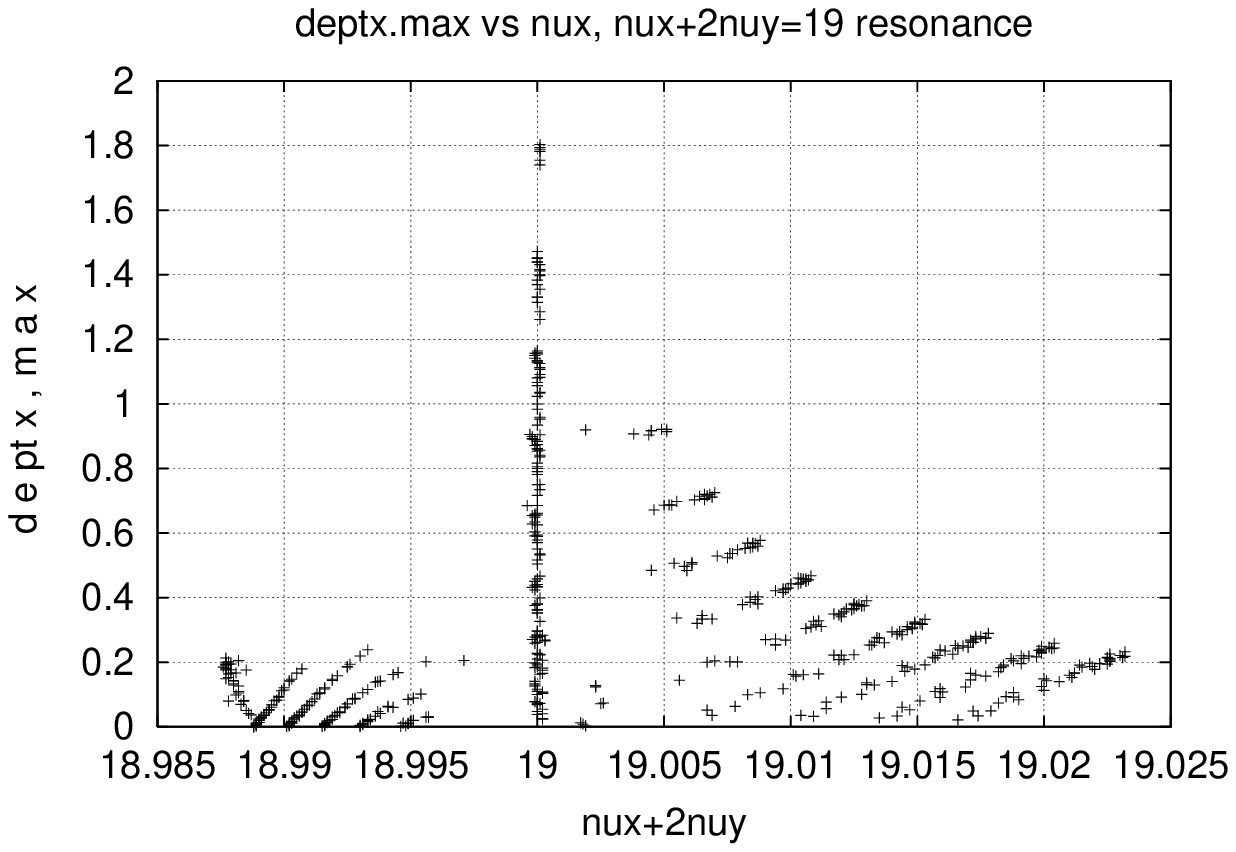,width=8.0cm}
\caption{$d\epsilon_{t,max}$ vs $\nu_x+2\nu_y$  
for the $\nu_x+2\nu_y=19$ resonance 
excited by a random $b_2$
in the SNS magnets . 
The initial coordinates $x_0,p_{x0},y_0,p_{y0}$ lie along 6 directions in
the $x_0,p_{x0}$ phase space and 6 directions in $y_0,p_{y0}$ space .$\nu_{x0}=6.3933$,$\nu_{y0}=6.2933$.
In the figure x0,px0,y0,py0,epx0,ept0,nux0,nuy0 represent
$x_0,p_{x0},y_0,p_{y0},\epsilon_{x0},\epsilon_{y0},\epsilon_{t0},
\nu_{x0},\nu_{y0}$.}
\label{fig3-6b}
\end{figure}

An interesting plot that may corrrespond more closely to something that might 
be measured is to plot the emittance growth, $d\epsilon_{t,max}$ vs  
$\nu_x+2\nu_y$. This plot can be found by combining 
Fig.~\ref{fig3-2} and Fig.~\ref{fig3-3} and is shown 
in Fig.~\ref{fig3-3b}. In this plot, there is a peak in 
the emittance growth which occurs when
$\nu_x+2\nu_y=19$. The peak could be used as a measure of the 
resonance strength to correct the resonance.


As in the 2-dimesional case, to correct the $\nu_x+2\nu_y=19$ resonance , 
one has to repeat the above calculations for all directions in phase space that
pass through a particular island, in order not to be misled by the
 movement of the islands when the correctors are changed. These results are 
shown in Fig.~\ref{fig3-4}, Fig.~\ref{fig3-5}.and Fig.~\ref{fig3-6} where 
for each $\epsilon_{t0}$ runs are done for
6 directions in $x_0,p_{x0}$ and for 6 directions in $y_0,p_{y0}$. In these figures there are,
then, 36 points for each value of $\epsilon_{t0}$. The results indicated by these figures 
for the width of the island and the maximun emittance growth 
are not too different from
those found above where only one direction in phase space was used, as 
the particular direction used appears to pass close to the center of the island.

An interesting plot that may corrrespond more closely to something that might 
be measured is to plot the emittance growth , $d\epsilon_{t,max}$ vs  $\nu_x+2\nu_y$. This plot can be found by combining Fig.~\ref{fig3-5} and Fig.~\ref{fig3-6} and is shown 
in Fig.~\ref{fig3-6b}. In this plot, there is a peak in the emittance growth which occurs when
$\nu_x+2\nu_y=19$. The peak could be used as a measure of the 
resonance strength to correct the resonance.

\begin{figure}[p]
\centering
\epsfig{file=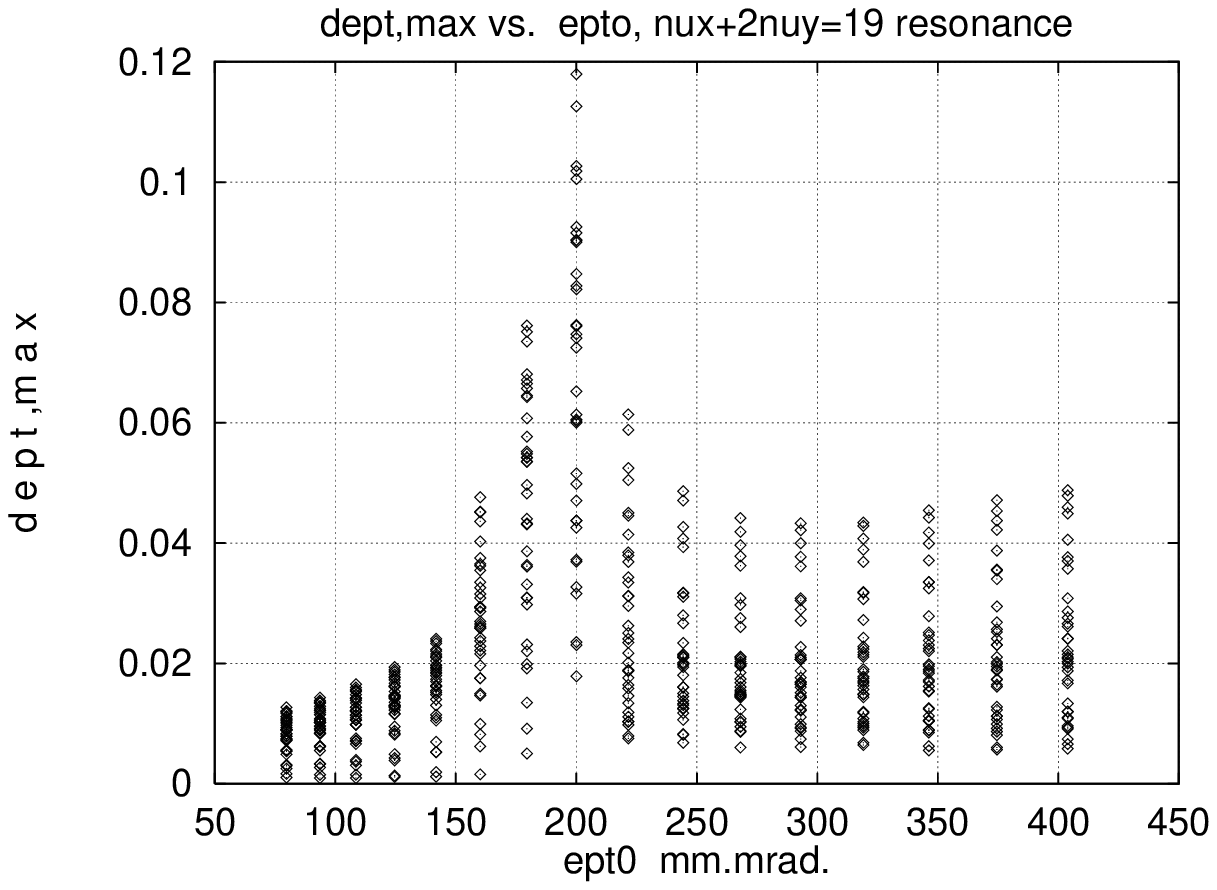,width=8.0cm}
\caption{The emittance sgrowth $d\epsilon_{t,max}$  vs $\epsilon_{t0}$  for the $\nu_x+2\nu_y=19$ resonance excited by a random $b_2$
in the SNS magnets and corrected using two sextupole correctors.
The initial coordinates $x_0,p_{x0},y_0,p_{y0}$ lie along 6 directions in
the x0, px0 phase space and 6 directions in $y_0,p_{y0}$ space .$\nu_{x0}=6.3933$,$\nu_{y0}=6.2933$.
In the figure x0,px0,y0,py0,epx0,ept0,nux0,nuy0 represent
$x_0,p_{x0},y_0,p_{y0},\epsilon_{x0},\epsilon_{y0},\epsilon_{t0},
\nu_{x0},\nu_{y0}$.} 
\label{fig3-7}
\end{figure}

\begin{figure}[p]
\centering
\epsfig{file=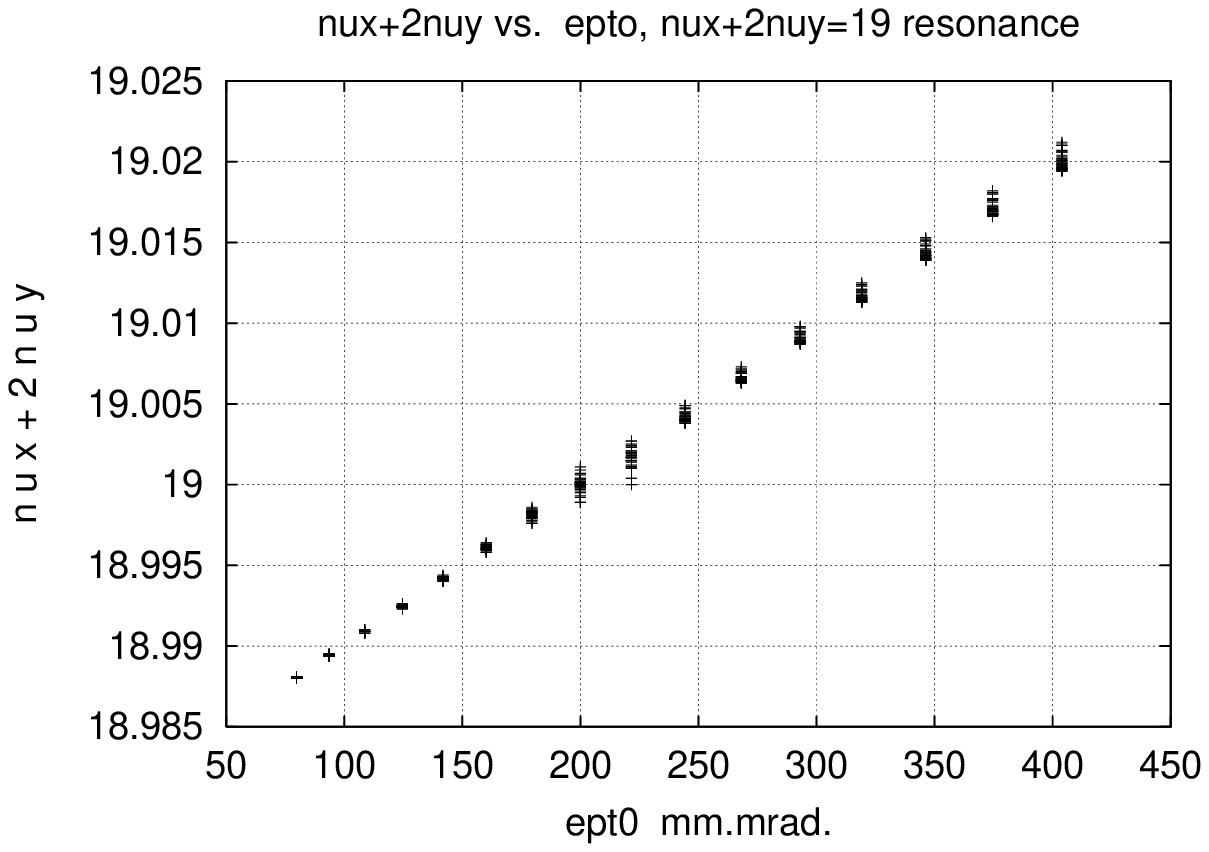,width=8.0cm}
\caption{$\nu_x+2|nu_y$  vs $\epsilon_{t0}$  for the $\nu_x+2\nu_y=19$ 
resonance excited by a random $b_2$
in the SNS magnets and corrected using two sextupole correctors.
The initial coordinates $x_0,p_{x0},y_0,p_{y0}$ lie along 6 directions in
the $x_0,p_{x0}$ phase space and 6 directions in $y_0,p_{y0}$ 
space .$\nu_{x0}=6.3933$,$\nu_{y0}=6.2933$.
In the figure x0,px0,y0,py0,epx0,ept0,nux0,nuy0 represent
$x_0,p_{x0},y_0,p_{y0},\epsilon_{x0},\epsilon_{y0},\epsilon_{t0},
\nu_{x0},\nu_{y0}$.} 
\label{fig3-8}
\end{figure}

The correction of the island resonances for the $\nu_x+2\nu_y=19$ 
resonance can be done 
with two sextupole correctors properly located around the ring. Assuming that one has a way to measure either the width of the island or the maximun emittance growth 
using the above results as a guide , then the two correctors 
can be adjusted one at a time to reduce the emittance growth or the island width.
Simulating 
this procedure gives the results shown in Fig.~\ref{fig3-7} and Fig.~\ref{fig3-8}.
The emittance growth ,the maximun $d\epsilon_{t,max}$, is reduced by almost a factor of 15 by 
the best setting of the correctors that was found. In this case, setting 
the correctors to zero the stopband, $d\nu_{12}$=0, reduces the maximun emittance
growth by a factor of 5.

\section*{Higher order resonances}

In the above, results were presented for the 3nux=19 and the $\nu_x+2\nu_y=19$ 
resonances. Similar studies have also been done for all four of the third order
resonances $m \nu_x+n \nu_y=19$, $ m+n=3$, and for all five of the fourth order resonances
$m \nu_x+n \nu_y=25$, $ m+n=4$, with similar results. It is our expectation 
that similar results
would be found for the higher order resonance $m \nu_x+n \nu_y=q$, where m,n,q are
integers. In particular, if $m \nu_x+n \nu_y$ is plotted versus $\epsilon_{t0}$, this plot will 
contain a flat region where  $m \nu_x+n \nu_y$ is costant 
at $m \nu_x+n \nu_y=q$, and 
the width
of this flat region will give the width of the islands associated with the
$m \nu_x+n \nu_y=q$ resonance.

\section*{ACKNOWLEDGEMENTS}

We would like to thank A.V. Fedotov for many discussions.


\begin{thebibliography}{99}
\bibitem{ypda}Y. Papaphilippou and D.T. Abell, EPAC'00, p.1453. (2000)
\bibitem{VPAFFP}V. Ptitsyn, A.V. Fedotov, F. Pilat,EPAC'02, p.350,(2002)
\bibitem{AFGP}A.V. Fedotov and G.Parzen, PAC'03, (2003)
\bibitem{GG}G.Guignard, CERN 77-10,(1977)
\end{thebibliography}
\end{document}